\documentclass{article}

\usepackage{lmodern}
\usepackage{cite}
\usepackage{amsmath,amssymb,amsfonts}
\usepackage{bm}
\usepackage{algorithmic}
\usepackage{graphicx,color}
\usepackage{textcomp}
\usepackage{booktabs}
\usepackage{pifont}   
\usepackage{hyperref}
\usepackage{cleveref}
\usepackage{xcolor}
\usepackage{multirow} 
\usepackage[margin=0.85in,top=0.72in,bottom=0.9in]{geometry}
\usepackage{fancyhdr}
\usepackage{titlesec}
\usepackage{lettrine}
\usepackage{newtxtext,newtxmath}
\usepackage{lmodern}  % 提供更好看的 Computer Modern
\renewcommand{\thesection}{\Roman{section}}
\renewcommand{\thesubsection}{\Alph{subsection}}
\titleformat{\section}[block]{\centering\normalfont\normalsize\scshape}{\thesection.}{0.6em}{}
\titleformat{\subsection}[runin]{\normalfont\itshape}{\thesubsection.}{0.5em}{}[.]
\titlespacing*{\section}{0pt}{1.0em}{0.7em}
\titlespacing*{\subsection}{0pt}{0.8em}{0.5em}
\begin{document}

%% ============================================================
%% Title
%% ============================================================

\twocolumn[
\begin{center}
{\fontfamily{lmr}\selectfont\LARGE\bfseries 
Phy2-ExposNet: A Physics-Informed Neural Network\\
for EMF Exposure Mapping in Complex Urban Environments
\par}

\vspace{0.8em}

{\normalsize Shuangning Li\textsuperscript{1}, Yarui Zhang\textsuperscript{2}, Shanshan Wang\textsuperscript{1}, and Joe Wiart\textsuperscript{1}\par}
\vspace{0.4em}

{\small

\textsuperscript{1}LTCI, Télécom Paris, Institut Polytechnique de Paris, Palaiseau, France. 

E-mail: \texttt{firstname.lastname@telecom-paris.fr}\\

\textsuperscript{2}SATIE, École normale supérieure Paris-Saclay, CNRS, Université Paris-Saclay, Gif-sur-Yvette, France. 

E-mail: \texttt{firstname.lastname@ens-paris-saclay.fr}

\par}
\vspace{1.0em}
\end{center}

]

\textbf{\textit{Abstract}}—
Accurate electromagnetic field (EMF) exposure mapping is critical for wireless network planning, environmental monitoring, and the deployment of next generation communication systems. The mapping results can be converted into the form of a radio map, a key technology in digital twin communication systems, used to describe the wireless signal propagation characteristics at every location in a specific area. Existing deep learning approaches treat propagation estimation as a pure regression problem and do not enforce physical consistency in the predicted fields. In this paper, we propose Phy2-ExposNet, a novel neural network framework that decouples exposure mapping into a physics-informed estimation stage and a transformer-based residual refinement stage. 
It first estimates the fields under two physical constraints and then refines the resulting exposure map by capturing long range interactions and complex spatial propagation patterns. 
Experiments demonstrate that the proposed method achieves lower estimation error while significantly reducing model complexity compared to existing approaches.
It achieves around 15\% relative error reduction over strong baselines, while using over 80\% fewer parameters than conventional physics-informed models. Ablation results further reveal that the physics-informed design is crucial for capturing complex propagation effects, particularly in boundary and shadow regions.

\textbf{\textit{Index Terms}}—electromagnetic field exposure, physics-informed neural networks, radio map

%% ============================================================
%% Introduction
%% ============================================================

\section{Introduction}
\label{sec:intro}

\lettrine[lines=2,lhang=0.08,nindent=0em]{E}{lectromagnetic} field (EMF) exposure estimation has become essential for modern wireless network planning and deployment.
As cellular systems continue to evolve toward 5G Advanced and beyond~\cite{emfSurvey2023, icnirp2020}, the demand for fast, scalable, and reliable radio propagation modeling is rapidly increasing.

EMF exposure maps are closely related to the broader concept of radio mapping, which aims to characterize the spatial distribution of wireless channel or signal-related quantities over a geographical area. Such maps, also referred to as radio maps or channel knowledge maps, have recently emerged as a key enabler for environment-aware communications in next-generation wireless systems, particularly in 6G networks~\cite{ckm2025}. By associating spatial locations with channel or signal information, radio maps support a wide range of applications, including beamforming, resource allocation, trajectory optimization, and intelligent network planning.

Among different types of radio maps, path loss maps and EMF exposure maps serve distinct but complementary purposes. Path loss maps describe the large-scale attenuation of wireless signals caused by propagation effects such as reflection, diffraction, and penetration, and are widely used for coverage prediction, base station deployment, and network optimization~\cite{pathloss2024}. In contrast, EMF exposure maps focus on the spatial distribution of electromagnetic field intensity, which is directly related to the level of radio-frequency energy absorbed by the human body or the environment. From a physical perspective, exposure is defined as the absorbed electromagnetic energy. In downlink scenarios, however, the variation due to the relative position and orientation between the human body and the device is typically negligible compared to the dominant effect of signal attenuation over distance and environment. Therefore, the exposure level can be reasonably approximated as being proportional to the received signal power. Under the commonly adopted single-source downlink assumption, this further implies that EMF exposure mapping is closely related to path loss estimation, establishing a practical connection between the two types of radio maps.

EMF exposure mapping has become increasingly important as it supports regulatory compliance, public health assessment, and risk perception management~\cite{emfSurvey2023, emfMapping2025}. 
In addition, it enables emerging applications such as energy harvesting and smart city monitoring, where accurate knowledge of electromagnetic field intensity is required for system design and optimization~\cite{emfMapping2025}. 

Despite decades of research, large-scale and high resolution EMF exposure mapping remains challenging.
In practice, the canonical approaches of exposure mapping include extensive measurement campaigns and measurement driven map reconstruction. 
Measurement based methods can provide reliable local exposure characterization, but they are labor intensive, costly to scale, and often require dense spatial sampling to obtain representative maps over large urban areas~\cite{guillen2025, exposnet2025}. 
To reduce this burden, surrogate and interpolation based approaches have also been investigated, where sparse measurements are combined with statistical or learning based models to reconstruct continuous exposure fields. 
Although such methods can improve efficiency, their accuracy still depends strongly on measurement coverage, sensor placement, and the ability of the surrogate model to generalize across heterogeneous environments~\cite{guillen2025, remsurvey2025, integratedPlatform2020}. 

Beyond measurement-based and interpolation methods, predictive approaches are generally classified into empirical models and geometry-based ray tracing methods. 
Empirical models, such as the Okumura Hata and COST 231 formulations, are computationally efficient but rely on scenario specific calibration and cannot adequately capture the complex spatial structure of propagation in dense urban environments~\cite{rsrpCNN2021}. 
Ray tracing methods, on the other hand, explicitly model wave interactions with surrounding structures and can achieve high accuracy~\cite{pmnet2024}. 
However, their computational cost grows rapidly with environmental complexity and operating frequency, which limits their practicality for real time inference and large scale map generation~\cite{airmap2025, pathlossChallenge2024}. 
As a result, there remains a clear need for methods that preserve the physical fidelity of propagation modeling while achieving the efficiency and scalability required for high resolution EMF exposure mapping.

In this context, deep learning based approaches have emerged as a promising solution for exposure map estimation. Levie et al.~\cite{radioUNet2021} introduced RadioUNet, demonstrating that convolutional neural networks can learn to estimate path loss with accuracy comparable to ray tracing simulations while achieving significantly lower inference cost. Following this, a number of CNN and encoder decoder based frameworks have been proposed for signal strength and reference signal received power (RSRP) prediction~\cite{rsrpCell2023, rsrpCNN2021, airmap2025, pmnet2024}. For example, AIRMap~\cite{airmap2025} achieves path gain prediction in just 4\,ms using a U-Net autoencoder operating on elevation maps, while PMNet~\cite{pmnet2024} further demonstrates strong generalization across unseen scenarios via transfer learning. These advances have also been extended to EMF exposure mapping, as illustrated by ExposNet~\cite{exposnet2025} and related frameworks targeting compliance monitoring in urban environments.

Nevertheless, purely data-driven approaches model propagation prediction as a regression problem without explicitly incorporating the underlying electromagnetic principles. In the absence of physical constraints, the learned models may produce predictions that violate physical laws of electromagnetic wave propagation. As a result, such models tend to overfit to the statistical patterns of the training data and exhibit limited generalization to unseen propagation scenarios.

To mitigate this issue, physics-informed machine learning has emerged as a promising paradigm by incorporating prior knowledge of electromagnetic propagation into the learning process. The seminal work of Raissi \emph{et al.}~\cite{raissi2019pinn} introduced physics-informed neural networks (PINNs), demonstrating that embedding partial differential equations (PDEs) into the training objective acts as an effective regularizer and improves data efficiency. In the electromagnetic domain, this idea has been extended to Maxwell's equations and their frequency-domain counterpart, the Helmholtz equation, enabling accurate field prediction and forward scattering modeling~\cite{pinnEM2024, pinnFEM2025}. In addition, Guo \emph{et al.}~\cite{physicsEmbeddedVIE2022} incorporated the volume integral equation (VIE) into deep architectures through algorithm unrolling, while Jiang \emph{et al.}~\cite{pefnet2024} directly embedded the VIE into the loss function for path loss prediction, establishing a hybrid physics and data-driven framework. Related efforts have also explored integrating domain knowledge into exposure map reconstruction, showing consistent improvements in accuracy and robustness~\cite{physicsInspiredRME2024, phyrmdm2025}.

Despite these advances, existing physics-informed approaches still exhibit important limitations. First, most methods rely on a single type of physical constraint, which provides only a partial description of electromagnetic propagation and may not fully capture the complex interactions present in realistic environments. Second, the majority of current models are built upon CNNs, which inherently rely on local receptive fields and hierarchical feature aggregation. While effective for modeling spatially localized patterns, CNNs struggle to capture long-range dependencies and global interactions, which are essential for accurately describing electromagnetic wave propagation involving reflection, diffraction, and scattering across large spatial regions. 

In contrast, transformer architectures~\cite{transformer2017}, have recently demonstrated strong capability in capturing long-range dependencies through self-attention mechanisms. By modeling global interactions across the entire spatial domain, transformers are well suited for representing the non-local characteristics of electromagnetic propagation. These properties make them particularly attractive for high-resolution exposure map estimation, where long-range wave interactions play a critical role.

Motivated by the above observations, we propose Phy2-ExposNet, a novel end-to-end framework for accurate urban EMF exposure estimation. The proposed method integrates two complementary physical constraints with a transformer-based refinement architecture within a coarse-to-fine modeling paradigm. Specifically, the first stage captures the global propagation structure under physics-informed supervision, while the second stage leverages the strong global modeling capability of transformers to refine spatial details and correct residual errors. By combining two forms of physical knowledge with long-range dependency modeling, the proposed approach overcomes the limitations of existing CNN-based and single-physics methods, leading to improved accuracy and generalization in complex propagation environments.

The main contributions of this work are summarized as follows:
\begin{itemize}

\item A physics-informed learning framework is proposed, which integrates two complementary physical constraints, namely the Helmholtz PDE and the volume integral equation (VIE), into a unified training objective. This composite loss design enforces electromagnetic consistency in the predicted complex field without requiring exact analytical solutions.

\item Develop a hybrid two-stage architecture that combines physics-informed field estimation with transformer-based refinement. The first stage leverages convolutional networks to capture structured propagation patterns and provide a physically meaningful initialization, while the second stage employs a transformer-based module to model long-range interactions and complex spatial dependencies beyond the capability of purely convolutional designs.

\item Experiments on two representative radio datasets demonstrate that Phy2-ExposNet consistently outperforms existing state-of-the-art methods, with notable improvements in challenging regions such as building boundaries and shadow areas.

\item Lightweight and ultra-lightweight variants are further introduced by reducing the complexity of the first-stage network. These variants significantly decrease the parameter count and memory footprint while maintaining competitive accuracy, highlighting the practicality of the proposed approach for resource-constrained deployment.

\end{itemize}
The remainder of this paper is organized as follows. Section~\ref{sec:background} introduces the 
electromagnetic background and theoretical formulations underlying the EMF Exposure. 
Section~\ref{sec:method} describes the Phy2-ExposNet network architecture in detail. 
Section~\ref{sec:experiments} presents simulation setup, datasets, and results. 
Section~\ref{sec:conclusion} concludes the paper.

%% ============================================================
%% Background
%% ============================================================

\section{Electromagnetic Background and Theoretical Formulations}
\label{sec:background}

This section establishes the theoretical fundamentals that the physics-informed training objectives used in Phy2-ExposNet. We introduce the relevant electromagnetic wave equations, formulate the PDE and VIE residual losses and derive the 2-D volume integral equation, and that are incorporated into the training procedure.

\subsection{Maxwell's Equations in the Frequency Domain}

In electromagnetic scattering, exposure assessment, and wireless propagation analysis, a time harmonic excitation is commonly assumed. Under this assumption, all field quantities vary as $e^{j\omega t}$, and Maxwell's equations can be expressed in the frequency domain as
\begin{equation}
\begin{aligned}
\nabla \times \mathbf{E} &= -j\omega \mathbf{B}, \\
\nabla \times \mathbf{H} &= \mathbf{J} + j\omega \mathbf{D}, \\
\nabla \cdot \mathbf{D} &= \rho, \\
\nabla \cdot \mathbf{B} &= 0.
\end{aligned}
\label{eq:maxwell}
\end{equation}
Here, $\mathbf{E}$ and $\mathbf{H}$ denote the electric and magnetic fields, $\mathbf{D}$ and $\mathbf{B}$ represent the electric flux density and magnetic flux density, $\mathbf{J}$ is the electric current density, $\rho$ is the charge density, and $\omega = 2\pi f$ is the angular frequency with $f$ being the operating frequency.

Equation ~\eqref{eq:maxwell} describe the fundamental coupling and conservation laws governing electromagnetic fields, independent of material properties. However, Maxwell's equations alone do not form a closed system, as they involve more unknowns than equations. To close the system, constitutive relations that characterize the material response are required. For linear, isotropic, and homogeneous media, these relations are given by
\begin{equation}
\mathbf{D} = \varepsilon \mathbf{E}, \quad
\mathbf{B} = \mu \mathbf{H}, \quad
\mathbf{J} = \sigma \mathbf{E},
\end{equation}
where $\varepsilon$ is the permittivity, $\mu$ is the permeability, and $\sigma$ is the electrical conductivity of the medium.

In the frequency domain, conduction and dielectric effects can be unified through the complex permittivity, which varies over spatial coordinates $(x,y)$ on the discretized exposure map grid:

\begin{equation}
  \varepsilon_c(x,y) = \varepsilon_0 \varepsilon_r(x,y) - j\frac{\sigma(x,y)}{\omega},
  \label{eq:permittivity}
\end{equation}

where $\varepsilon_0$ is the permittivity of free space, $\varepsilon_r(x,y)$ denotes the relative permittivity of the medium, and $\sigma(x,y)$ is the spatially varying electrical conductivity. The real part of $\varepsilon_c$ represents the energy storage capability of the medium, while the imaginary part accounts for conductive and dielectric losses. This formulation provides a unified description of wave attenuation and dispersion in heterogeneous environments.

\subsection{Helmholtz Equation Formulation}

The frequency domain Maxwell equations together with the constitutive relations form a complete physical description of electromagnetic fields. However, they involve both the electric field $\mathbf{E}$ and the magnetic field $\mathbf{H}$.  By taking the curl of Faraday's law in~\eqref{eq:maxwell} and using Amp\`ere's law, the system can be reformulated in terms of the electric field $\mathbf{E}$ alone:
%To this end, the magnetic field $\mathbf{H}$ can be eliminated \sw{by replacing ...}. Specifically, taking the curl of Faraday's law in~\eqref{eq:faraday} and substituting Amp\`ere's law yields the electric field wave equation
\begin{equation}
\nabla \times \nabla \times \mathbf{E}
- \omega^2 \mu \varepsilon_c \mathbf{E}
=
-j\omega \mu \mathbf{J},
\label{eq:wave_equation_general}
\end{equation}
where $\varepsilon_c$ is the complex permittivity defined in~\eqref{eq:permittivity}. This equation is commonly called as the frequency domain wave equation and is directly derived from Maxwell's equations.

Equation~\eqref{eq:wave_equation_general} describes the propagation, scattering, and attenuation of electromagnetic waves in heterogeneous media. The term $\nabla \times \nabla \times \mathbf{E}$ captures spatial field variations, while the term $\omega^2 \mu \varepsilon_c \mathbf{E}$ represents the combined effects of wave propagation and material response.

To further simplify the formulation, we use the vector identity
\begin{equation}
\nabla \times \nabla \times \mathbf{E}
=
\nabla(\nabla \cdot \mathbf{E}) - \nabla^2 \mathbf{E}.
\end{equation}

In source-free regions with negligible free charge accumulation, i.e., $\nabla \cdot (\varepsilon \mathbf{E}) \approx 0$, the divergence term becomes negligible. Under this assumption, the wave equation in~\eqref{eq:wave_equation_general} reduces to the vector Helmholtz equation
\begin{equation}
\nabla^2 \mathbf{E} + k^2 \mathbf{E} = 0,
\label{eq:vector_helmholtz_final}
\end{equation}
where $\nabla^2$ is Laplacian operator and the complex wavenumber is given by $k^2 = \omega^2 \mu \varepsilon_c$.

The wavenumber $k$ encodes both phase propagation and attenuation Its real part determines the wavelength and phase velocity, while its imaginary part accounts for energy dissipation due to material losses.

Because the exact permittivity distribution is spatially heterogeneous and typically unknown in practical scenarios, directly using the true wavenumber $k$ is infeasible. Instead, following the common practice in physics-informed neural networks of relaxing exact physical parameters into learnable or effective quantities~\cite{raissi2019pinn}, we introduce a uniform effective parameter $\beta \geq 0$ to approximate $k^2$. This leads to a relaxed PDE residual that enforces approximate propagation consistency rather than strict equation satisfaction. The Helmholtz equation is thus simplified as

\begin{equation}
  \nabla^2 \mathbf{E}_{\text{tot}} + \beta \, \mathbf{E}_{\text{tot}} \approx 0.
  \label{eq:helmholtz_approx}
\end{equation}

Since the predicted field is complex valued with real and imaginary components 
%$\mathbf{E}^{\text{tot}} = \mathbf{E}_{\text{re}} + j\mathbf{E}_{\text{im}}$, 
$\mathbf{E}_{\text{tot}} = \Re(\mathbf{E}_{\text{tot}}) + j \Im(\mathbf{E}_{\text{tot}})$
applying \eqref{eq:helmholtz_approx} separately to each component yields:

\begin{equation}
\begin{aligned}
  \nabla^2 \Re(\mathbf{E}_{\text{tot}}) + \beta \Re(\mathbf{E}_{\text{tot}}) &\approx 0, \\
  \nabla^2 \Im(\mathbf{E}_{\text{tot}}) + \beta \Im(\mathbf{E}_{\text{tot}}) &\approx 0.
\end{aligned}
\label{eq:helm_complex_split}
\end{equation}

\subsection{Volume Integral Equation and Discretized Formulation}
\label{subsec:vie}

The volume integral equation (VIE) provides a global, scattering theoretic description  of electromagnetic propagation~\cite{physicsEmbeddedVIE2022, vieScattering2022}. 
For the 2-D transverse magnetic (TM) polarization, the total electric field at the observation point $\mathbf{p}$ satisfies
\iffalse
\begin{equation}
  E_{\text{tot}}(\mathbf{p}) = E_{\text{inc}}(\mathbf{p}) + k_0^2 \int_D G(\mathbf{p} - \mathbf{p}') \,
  \chi(\mathbf{p}') \, E_{\text{tot}}(\mathbf{p}') \, d\mathbf{p}',
  \label{eq:vie_integral}
\end{equation}
\fi
\begin{equation}
  E_{\mathrm{tot}}(\mathbf{p}) 
  = E_{\mathrm{inc}}(\mathbf{p}) 
  + k_0^2 \int_D 
  G(\mathbf{p} - \mathbf{p}') 
  \chi(\mathbf{p}') 
  E_{\mathrm{tot}}(\mathbf{p}') 
  \, \mathrm{d}\mathbf{p}' .
  \label{eq:vie_integral}
\end{equation}
where $E_{\text{inc}}(\mathbf{p}) = G(\mathbf{p}_{\text{tx}} - \mathbf{p})$ denotes the incident field generated by the transmitter located at $\mathbf{p}_{\text{tx}}$. The function $G(\mathbf{p})$ denotes the 2-D free space Green's function, and $\chi(\mathbf{p})$ is the contrast function defined as
\begin{equation}
  \chi(\mathbf{p}) = \frac{\varepsilon_c(\mathbf{p}) - \varepsilon_0}{\varepsilon_0},
  \label{eq:contrast}
\end{equation}
which characterizes the relative deviation of the local medium from free space. From this definition, $\chi(\mathbf{p}) = 0$ in free space where $\varepsilon_c = \varepsilon_0$, while nonzero values of $\chi(\mathbf{p})$ arise in building regions due to material contrast.

The 2-D free space Green's function for TM polarization is denoted as:

\begin{equation}
  G(\mathbf{p} - \mathbf{p}') = -\frac{j}{4} H_0^{(2)}\!\left(k_0 |\mathbf{p} - \mathbf{p}'|\right),
  \label{eq:greens}
\end{equation}

where $H_0^{(2)}$ is the Hankel function of the second kind of order zero.
To enable numerical implementation and efficient computation, the continuous integral equation must be discretized over the spatial domain. Discretizing the 2-D domain into $N = N_x \times N_y$ cells of equal area $A$ and 
approximating each cell as a circular equivalent disk of radius $a = \sqrt{A/\pi}$, 
the continuous integral equation~\eqref{eq:vie_integral} takes the matrix form:

\begin{equation}
  (\mathbf{I} + \mathbf{W}\boldsymbol{\chi}) \, \mathbf{E}_{\text{tot}} = \mathbf{E}_{\text{inc}},
  \label{eq:vie_matrix}
\end{equation}

where $\mathbf{E}^{\text{tot}}, \mathbf{E}^{\text{inc}} \in \mathbb{C}^{N\times 1}$ are the 
vectorized total and incident field values, $\boldsymbol{\chi} = \text{diag}(\chi_1, \ldots, \chi_N)$ 
is the diagonal contrast matrix, and $\mathbf{W} \in \mathbb{C}^{N\times N}$ is the discretized 
Green's function operator with entries:

\begin{equation}
  W_{n,s} = \begin{cases}
    \dfrac{j}{2}\!\left[\pi k_0 a \, H_1^{(2)}(k_0 a) - 2j\right] & n = s,\\[8pt]
    \dfrac{j\pi k_0 a}{2} J_1(k_0 a) \, H_0^{(2)}\!\left(k_0 |\mathbf{p}_n - \mathbf{p}_s|\right) & n \neq s,
  \end{cases}
  \label{eq:green_matrix}
\end{equation}

where $J_1$ and $H_1^{(2)}$ are the Bessel and Hankel functions of order one, respectively. 
The diagonal self interaction term accounts for the singular Green's function at $\mathbf{p}_n = \mathbf{p}_s$ 
via the standard equivalent disk regularization.

%% ============================================================
%% Network Architecture
%% ============================================================

\section{Phy2-ExposNet Framework}
\label{sec:method}

This section presents the overall framework of the proposed network. The framework follows a two-stage design, consisting of physics-informed field estimation and transformer-based residual refinement. The first stage estimates a physically consistent electromagnetic field from the input representation, while the second stage refines the corresponding field by learning residual corrections.

\begin{figure*}[t]
    \centering
    \includegraphics[width=0.8\textwidth]{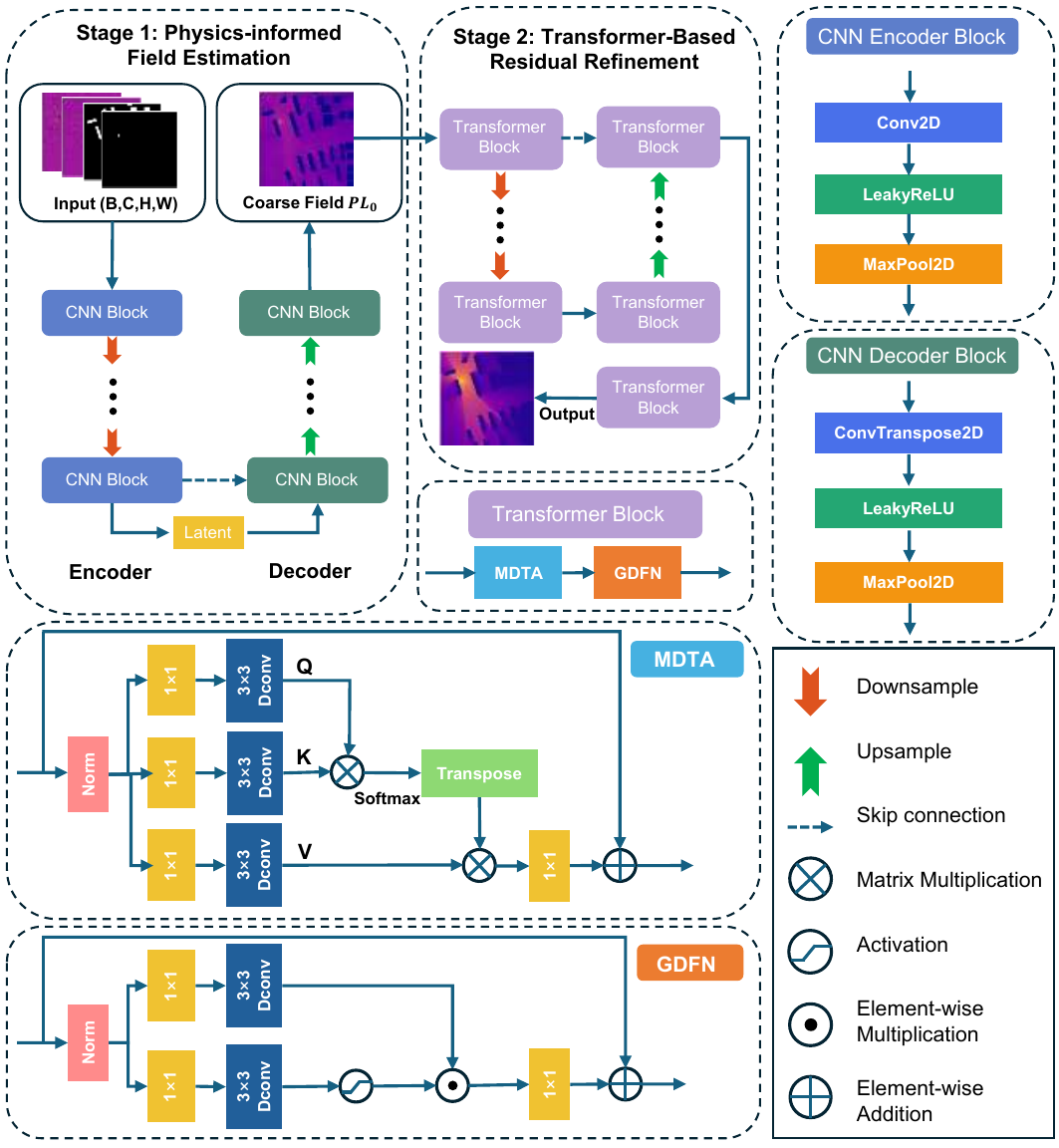}
    \caption{Overview of the proposed two stage architecture for EMF exposure estimation.}
    \label{fig:network}
\end{figure*}

\subsection{Stage 1: Physics-Informed Field Estimation}

The first stage takes as input a multi-channel tensor that encodes both environmental geometry and physical priors. Specifically, the input is constructed as
\begin{equation}
\mathbf{X} = \left[ \mathbf{M}_{\text{bld}}, \mathbf{M}_{\text{tx}}, \Re(\mathbf{E}_{\text{inc}}), \Im(\mathbf{E}_{\text{inc}}) \right],
\end{equation}
where each component is represented as a 2-D grid of size $H \times W$.

The spatial domain is uniformly discretized into $N = H \times W$ cells, where each pixel corresponds to a spatial location in the propagation environment. The building mask $\mathbf{M}_{\text{bld}} \in \{0,1\}^{H \times W}$ is a binary map indicating the presence of obstacles, with value 1 for grid cells occupied by buildings and 0 otherwise. Similarly, the transmitter mask $\mathbf{M}_{\text{tx}} \in \{0,1\}^{H \times W}$ encodes the transmitter location, where the grid containing the transmitter is set to 1 and all other locations are set to 0.

To incorporate physical prior knowledge, the incident electric field $\mathbf{E}_{\text{inc}}$ is pre-computed through Green function from Eq. \ref{eq:vie_integral}. Since $\mathbf{E}_{\text{inc}}$ is complex-valued, it is decomposed into its real and imaginary components, i.e., $\Re(\mathbf{E}_{\text{inc}})$ and $\Im(\mathbf{E}_{\text{inc}})$, which are treated as two separate input channels. Each component is computed at the grid level and normalized to a comparable scale with the other input features.

All input features are represented as single-channel images and concatenated along the channel dimension, resulting in a four-channel input tensor. 
The network estimates the initialed field as:

\begin{equation}
\hat{\mathbf{E}}_{\text{tot}} = \mathcal{F}_{\theta_1}(\mathbf{X}).
\end{equation}

The inclusion of the incident field $\mathbf{E}_{\text{inc}}$ provides an explicit physical prior, allowing the network to focus on learning the scattering effects induced by the environment. This reduces the learning difficulty and improves physical consistency.

The mapping function $\mathcal{F}_{\theta_1}(\cdot)$ is implemented using a lightweight encoder-decoder architecture with skip connections. As illustrated in Fig.~\ref{fig:network}, both the encoder and decoder consist of seven convolutional blocks, enabling hierarchical feature extraction and reconstruction. The encoder captures large-scale propagation patterns such as diffraction and shadowing, while the decoder reconstructs fine spatial details by integrating multi-scale features.

An initial path loss map is then derived from the predicted coarse Field as shown in Fig.~\ref{fig:network}:
 
\begin{equation}
\mathbf{PL}_0 = \mathcal{G}(\hat{\mathbf{E}}_{\text{tot}}),
\end{equation}
where $\mathcal{G}(\cdot)$ converts the complex field into a path loss representation by computing its magnitude and mapping it to a logarithmic scale. This transformation reflects the inverse relationship between field strength and path loss in wireless propagation.

By embedding physical priors into the input representation, this stage produces a structured and physically meaningful estimate, which serves as a reliable initialization for further refinement.

\subsection{Stage 2: Transformer-Based Residual Refinement}

To improve the prediction accuracy, a refinement module is introduced to estimate the residual error $\Delta \mathbf{PL}$, and the final output is $\hat{\mathbf{PL}}$:

\begin{equation}
\Delta \mathbf{PL} = \mathcal{F}_{\theta_2}(\mathbf{PL}_0), \quad
\hat{\mathbf{PL}} = \mathbf{PL}_0 + \Delta \mathbf{PL}.
\end{equation}

The refinement function $\mathcal{F}_{\theta_2}(\cdot)$ adopts a lightweight transformer-based encoder–decoder architecture, composed of multi-Dconv head transposed attention (MDTA) and gated depthwise feed-forward network (GDFN) blocks. As illustrated in Fig.~\ref{fig:network}, the refinement network consists of seven Transformer blocks, enabling progressive feature refinement from coarse to fine scales. This design enables efficient modeling of both global dependencies and local variations in electromagnetic propagation.

Unlike conventional self-attention mechanisms that operate over spatial dimensions, where each spatial location attends to all others~\cite{transformer2017}, the MDTA module performs attention along the channel dimension. Specifically, query, key, and value representations are first generated through pointwise projections, followed by depthwise convolution to incorporate local spatial context. The attention is then computed across channels, resulting in a compact attention matrix that captures inter-channel correlations rather than pairwise spatial interactions.

This channel-oriented attention significantly reduces computational complexity from quadratic scaling with respect to spatial resolution, while still enabling global information exchange. By modeling dependencies between feature channels, the network can effectively capture different propagation responses and their interactions, which is essential for representing complex electromagnetic phenomena such as multipath propagation and long-range coupling.

To further enhance feature representation, each attention block is followed by a gated depthwise feed-forward network (GDFN). In this module, features are first expanded and processed using depthwise convolution, then split into two branches, where one branch modulates the other via a gating mechanism~\cite{chai2020highway}. This adaptive feature selection improves the network’s ability to capture fine-grained variations, particularly near building boundaries and shadow transition regions.

The refinement stage follows a residual learning strategy. Given the coarse prediction $\mathbf{PL}_0$ from Stage 1, the network learns a residual correction $\Delta \mathbf{PL}$ to produce the final estimate. During training, supervision is applied to the final refined output, while the intermediate output $\mathbf{PL}_0$ serves as a physically meaningful initialization that guides the learning process.

\subsection{Overall Framework}

The overall mapping separates the learning process into two complementary tasks. The first stage captures the global propagation structure under physical guidance, while the second stage refines local details and corrects residual errors.

Such a decomposition improves both interpretability and performance. The first stage introduces a strong physical prior without requiring explicit solution of the full wave equation, and the second stage enhances spatial accuracy by focusing on challenging regions. As a result, the proposed framework achieves a balanced trade off between physical consistency, prediction accuracy, and computational efficiency.

\subsection{Loss function design}
This section presents a physics-informed loss design that combines two complementary physical constraints, namely PDE and VIE residual losses, with data-driven supervision to jointly train the network.

\subsubsection{PDE Residual Loss}
\label{subsec:pde_loss}

Given the decomposition above, the PDE residual for the predicted field $\hat{\mathbf{E}}_{\text{tot}}$ is defined as:

\begin{equation}
    R_{\text{PDE}} =
    \left( \nabla^2 \Re(\hat{\mathbf{E}}_{\text{tot}}) - \beta \Re(\hat{\mathbf{E}}_{\text{tot}}) \right)^2
    +
    \left( \nabla^2 \Im(\hat{\mathbf{E}}_{\text{tot}}) - \beta \Im(\hat{\mathbf{E}}_{\text{tot}}) \right)^2.
    \label{eq:pde_residual}
\end{equation}

The Laplacian operator $\nabla^2$ is approximated numerically on the discrete prediction grid 
using the standard five-point finite difference stencil:

\begin{equation}
\begin{aligned}
\nabla^2 f(i,j) \approx {} & f(i+1,j) + f(i-1,j) \\
& + f(i,j+1) + f(i,j-1) - 4f(i,j)
\end{aligned}
\label{eq:laplacian_fd}
\end{equation}

which is implemented efficiently as a fixed $3\times3$ convolutional kernel. 
The corresponding PDE training loss is:

\begin{equation}
  \mathcal{L}_{\text{PDE}} = \frac{1}{N} \sum_{x,y} R_{\text{PDE}},
  \label{eq:loss_pde}
\end{equation}

where $N = H \times W$ is the number of spatial grid points.

\subsubsection{VIE Residual Loss}
\label{subsec:vie_loss}

Given the neural network's predicted complex total field, 
the VIE residual measures the extent to which the prediction satisfies the physical coupling between the incident field, the scatterer, and the total field:

\begin{equation}
  \mathbf{R}_{\text{VIE}} = (\mathbf{I} + \mathbf{W}\boldsymbol{\chi})\hat{\mathbf{E}}_{\text{tot}} - \mathbf{E}_{\text{inc}}.
  \label{eq:vie_residual}
\end{equation}

Then, the VIE training loss is:

\begin{equation}
  \mathcal{L}_{\text{VIE}} = \frac{1}{N} \left\| (\mathbf{I} + \mathbf{W}\boldsymbol{\chi})\hat{\mathbf{E}}_{\text{tot}} - \mathbf{E}_{\text{inc}} \right\|_2^2.
  \label{eq:loss_vie}
\end{equation}

Because $\mathbf{W}$ is a dense and shift-invariant operator, its entries depend only on the relative distance $|\mathbf{p}_n - \mathbf{p}_s|$ between spatial locations. This property allows the matrix–vector product involving $\mathbf{W}$ to be reformulated as a two-dimensional convolution with a distance-dependent kernel. 

However, directly performing this operation in matrix form incurs a computational cost of $\mathcal{O}(N^2)$, which is prohibitive for high-resolution maps. To address this issue, we implement the convolution efficiently in the Fourier domain using the Fast Fourier Transform (FFT), reducing the complexity to $\mathcal{O}(N \log N)$. 

This formulation significantly improves scalability and enables practical implementation of the VIE operator within the learning framework.

\subsubsection{Composite Physics Loss}
\label{subsec:composite_loss}

The two physics constraints from Eq. \ref{eq:loss_pde} and \ref{eq:loss_vie} are combined into a composite physics-informed loss for network training:

\begin{equation}
  \mathcal{L}_{\text{phys}} = \lambda_{\text{PDE}} \mathcal{L}_{\text{PDE}} + \lambda_{\text{VIE}} \mathcal{L}_{\text{VIE}},
  \label{eq:loss_phys}
\end{equation}

where $\lambda_{\text{PDE}}$ and $\lambda_{\text{VIE}}$ are weighting hyperparameters. 
The total training loss combines the data driven supervised term with this physics knowledge:

\begin{equation}
  \mathcal{L} = \mathcal{L}_{\text{data}} + \mathcal{L}_{\text{phys}},
  \label{eq:loss_total}
\end{equation}

where $\mathcal{L}_{\text{data}} = \|\hat{y} - y\|_2^2$ is the mean squared error between the predicted and ground-truth exposure maps. The PDE loss enforces local propagation consistency by penalizing violations of the Helmholtz equation in individual cells, while the VIE loss enforces a global incident to total field coupling constraint that captures the collective scattering effect of all building elements.
Together, these two terms act as complementary physical constraints, guiding the network toward electromagnetically consistent predictions without requiring exact Maxwell solutions.

%% ============================================================
%% Simulation and results
%% ============================================================

\section{Simulation Setup and Results}
\label{sec:experiments}

This section presents the experimental setup and evaluation results of the proposed framework. We first describe the datasets, implementation details, and evaluation metrics, and then provide both quantitative and qualitative comparisons with existing methods. 

\subsection{Dataset}
\label{subsec:dataset}

We evaluate the Phy2-ExposNet on two representative datasets covering both simulated and real-world propagation conditions, namely RadioMapSeer~\cite{radioUNet2021} and  USC-PL dataset~\cite{pmnet2024}. The details of the dataset can be found in the Table~\ref{tab:dataset_stats}.

RadioMapSeer is a large-scale publicly available benchmark for urban path loss prediction. It consists of 700 scenarios with 80 transmitter locations per scenario, resulting in a total of 56{,}000 propagation instances covering diverse urban environments. The dataset provides simulated path loss maps at 5.9\,GHz on a $256 \times 256$ grid with a spatial resolution of 1\,m. The urban layouts are constructed from OpenStreetMap building footprints of Munich. Each sample includes a building occupancy map, a transmitter location mask, and the corresponding path loss map generated using a full wave electromagnetic simulator based on the Dominant Path Model (DPM). 

USC-PL is a dataset derived from the PMNet (Pathloss Map Network) framework~\cite{pmnet2024}, which is designed for data-driven wireless propagation modeling. It captures realistic wireless propagation characteristics in urban environments. The dataset is collected at a carrier frequency of 2.5\,GHz with an isotropic transmitter antenna and a transmit power of 0\,dBm.

The original maps cover an area of $880 \times 880$\,m$^2$, which are further cropped into smaller patches of size $221 \times 221$, corresponding to an effective spatial resolution of approximately 0.86\,m per pixel. The dataset contains a total of 4754 samples, each associated with a single transmitter configuration.

Compared to first dataset, USC-PL reflects real-world signal attenuation with measurement noise and environmental uncertainties, making it suitable for evaluating the robustness and generalization capability of network models.

For both datasets, we follow the standard train, valid and test splits commonly adopted in prior works\cite{radioUNet2021, pefnet2024}, where the data are partitioned into disjoint subsets for training, hyperparameter tuning, and evaluation to ensure fair comparison.

\begin{table*}[t]
\centering
\caption{Basic statistics of the adopted path loss datasets.}
\label{tab:dataset_stats}
\small
\setlength{\tabcolsep}{4pt}
\renewcommand{\arraystretch}{1.15}
\begin{tabular}{l|cccccccc}
\hline
\textbf{Dataset}
& \textbf{Frequency} 
& \textbf{Image size} 
& \textbf{Pixel length} 
& \textbf{Type} 
& \textbf{Features} 
& \textbf{Total samples} \\
\hline
RadioMapSeer 
& 5.9\,GHz 
& $256 \times 256$ 
& 1\,m 
& Simulation 
& Buildings 
& 56,000 \\
\hline
USC-PL 
& 2.5\,GHz 
& $221 \times 221$ 
& 0.86\,m 
& Simulation 
& Terrain + Buildings 
& 4754 \\
\hline
\end{tabular}
\end{table*}

\subsection{Implementation Details}
\label{subsec:implementation}

All models are implemented in PyTorch and trained on an NVIDIA A100 GPU cluster equipped with 40GB GPUs, AMD EPYC 7302 CPUs, and 384 GB system memory.

The model is trained for 50 epochs with a batch size of 16. 
The weights of the physics losses are set to $\lambda_{\text{VIE}} = 0.5$ and $\lambda_{\text{PDE}} = 0.5$, providing a balanced contribution between the VIE-based and PDE-based constraints. 

For the PDE-based regularization, the coefficient $\beta$ is set to 0.1 to control the strength of the Helmholtz-type constraint. 

The training strategy is formulated to integrate physics-informed learning with data-driven supervision in a unified manner. Specifically, the output of Stage 1 is used to compute the physics-informed losses, including the PDE and VIE residuals, while the output of Stage 2 is supervised by the data loss. These objectives are jointly optimized to train the entire network end-to-end.

This configuration allows the model to simultaneously leverage electromagnetic propagation principles and supervised learning signals, leading to stable training and improved generalization performance in complex propagation environments.

\subsection{Evaluation Metrics}
\label{subsec:metrics}

We evaluate the estimation performance using four complementary metrics: normalized mean square error (NMSE), root mean square error (RMSE), mean absolute error (MAE), and structural similarity index (SSIM). These metrics jointly assess global reconstruction accuracy, average deviation, and spatial structure preservation of the predicted EMF exposure maps.

Let $\mathbf{Y} \in \mathbb{R}^{H \times W}$ denote the ground-truth EMF exposure map and $\hat{\mathbf{Y}} \in \mathbb{R}^{H \times W}$ the predicted map. The value at pixel $(i,j)$ is denoted by $Y_{i,j}$ and $\hat{Y}_{i,j}$, respectively.

\subsubsection{Normalized Mean Square Error}
The NMSE is defined as
\begin{equation}
\mathrm{NMSE}
=
\frac{\sum_{i=1}^{H}\sum_{j=1}^{W}\left(\hat{Y}_{i,j}-Y_{i,j}\right)^2}
{\sum_{i=1}^{H}\sum_{j=1}^{W}Y_{i,j}^{2}}.
\label{eq:nmse}
\end{equation}
NMSE measures the squared prediction error normalized by the energy of the ground-truth map, and therefore reflects the overall relative reconstruction accuracy. A smaller NMSE indicates better performance. For quantitative comparison, NMSE is also reported in decibel scale as
\begin{equation}
\mathrm{NMSE}_{\mathrm{dB}} = 10 \log_{10}(\mathrm{NMSE}).
\label{eq:nmse_db}
\end{equation}

\subsubsection{Root Mean Square Error (RMSE)}
The RMSE is given by
\begin{equation}
\mathrm{RMSE}
=
\sqrt{
\frac{1}{HW}
\sum_{i=1}^{H}\sum_{j=1}^{W}
\left(\hat{Y}_{i,j}-Y_{i,j}\right)^2
}.
\label{eq:rmse}
\end{equation}
RMSE quantifies the average magnitude of the prediction error and is more sensitive to large deviations, providing an intuitive measure of absolute accuracy.

\subsubsection{Mean Absolute Error (MAE)}
The MAE is defined as
\begin{equation}
\mathrm{MAE}
=
\frac{1}{HW}
\sum_{i=1}^{H}\sum_{j=1}^{W}
\left|\hat{Y}_{i,j}-Y_{i,j}\right|.
\label{eq:mae}
\end{equation}
MAE measures the average absolute deviation between the predicted and ground-truth maps and is less sensitive to outliers compared to RMSE.

\subsubsection{Structural Similarity Index (SSIM)}
To evaluate spatial consistency, we adopt the structural similarity index (SSIM), defined as
\begin{equation}
\mathrm{SSIM}(\hat{\mathbf{Y}}, \mathbf{Y})
=
\frac{(2\mu_{\hat{Y}}\mu_Y + C_1)(2\sigma_{\hat{Y}Y} + C_2)}
{(\mu_{\hat{Y}}^2 + \mu_Y^2 + C_1)(\sigma_{\hat{Y}}^2 + \sigma_Y^2 + C_2)}.
\label{eq:ssim}
\end{equation}
where $\mu_{\hat{Y}}$ and $\mu_Y$ denote the mean values, $\sigma_{\hat{Y}}^2$ and $\sigma_Y^2$ the variances, and $\sigma_{\hat{Y}Y}$ the covariance. SSIM evaluates similarity in terms of luminance, contrast, and structural information, and is particularly important for preserving propagation patterns such as shadowing and blockage. A larger SSIM indicates higher structural similarity.

\subsection{Numerical Results}
\label{subsec:Numerical Results}

\subsubsection{Overall Performance}
\label{subsubsec:overall}
\begin{figure*}[ht]
    \centering
    \includegraphics[width=0.8\textwidth]{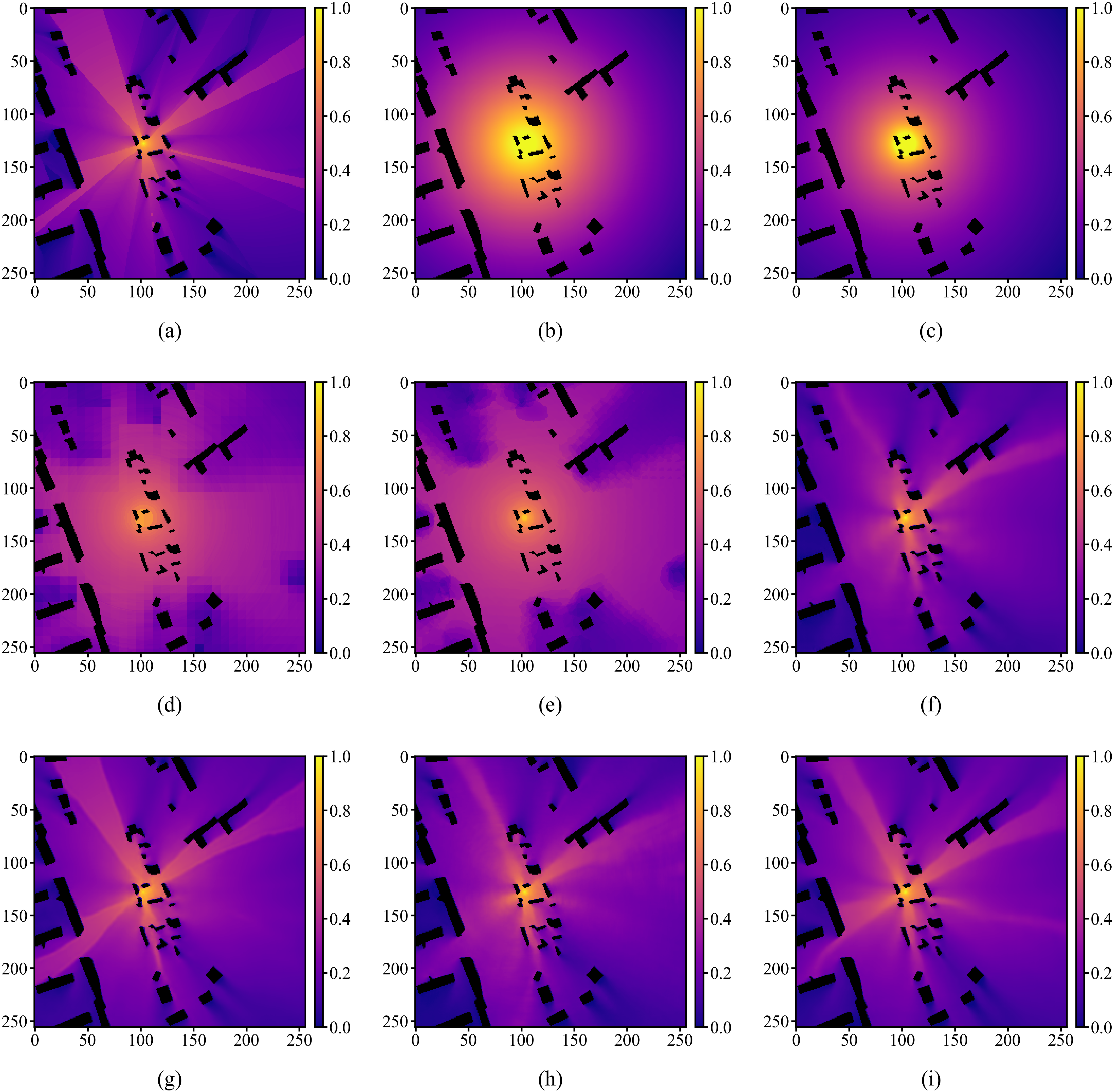}
    \caption{Visualization comparison of different methods on the RadioMapSeer dataset. The predicted values are normalized to the range $(0,1)$. From (a) to (i): (a) Ground truth, (b) ITU-R IMT-2020, (c) 3GPP 38.901, (d) XGBoost, (e) KNN, (f) RadioUNet, (g) PEFNet, (h) ResNet, (i) Phy2-ExposNet.}

    \label{fig:vis}
\end{figure*}

Fig.~\ref{fig:vis} presents a qualitative comparison of different methods on two dataset. The ground truth in Fig.~\ref{fig:vis}(a) exhibits clear propagation characteristics, including radial attenuation from the transmitter, strong shadowing behind buildings, and directional energy leakage along streets and open spaces. These patterns reflect the complex interaction between electromagnetic waves and the urban environment.
The empirical models in Fig.~\ref{fig:vis}(b) and Fig.~\ref{fig:vis}(c), corresponding to ITU-R IMT-2020 and 3GPP38.901, fail to capture the specific propagation behavior. Their predictions are overly smooth and mainly dominated by distance dependent attenuation, without properly reflecting blockage and diffraction effects caused by buildings. As a result, shadow regions and electromagnetic propagation patterns are largely missing.

The classical machine learning approaches, including XGBoost in Fig.~\ref{fig:vis}(d) and KNN in Fig.~\ref{fig:vis}(e), provide limited improvements. Although they partially incorporate environmental features, their predictions remain locally inconsistent and suffer from discontinuities. These methods lack an explicit mechanism to model spatial correlations, leading to blurred structures and inaccurate representation of propagation boundaries.

The deep learning baseline RadioUNet~\cite{radioUNet2021} in Fig.~\ref{fig:vis}(f) is a U-Net-based architecture that learns the mapping from environmental inputs to path loss maps using convolutional encoding and decoding. While it captures the overall attenuation trend, it produces over-smoothed predictions and fails to accurately reconstruct shadowing effects.

The PEFNet model \cite{pefnet2024} in Fig.~\ref{fig:vis}(g) extends this framework by incorporating physics-inspired constraints, specifically through the integration of the VIE into the training loss. This improves structural consistency to some extent, particularly around obstacles. However, it still struggles to preserve long-range propagation characteristics and exhibits noticeable deviations in complex regions.

The ResNet-based variant in Fig.~\ref{fig:vis}(h) adopts residual learning to enhance feature representation and improve local detail reconstruction. It is built upon stacked residual blocks with identity skip connections, which facilitate gradient propagation and enable deeper network design. Nevertheless, it does not fully recover directional propagation patterns and exhibits artifacts in regions with strong diffraction.

\begin{table*}[t]
\centering
\caption{Performance comparison on two datasets.}
\label{tab:perf_compare_two_datasets}
\small
\setlength{\tabcolsep}{4pt}
\renewcommand{\arraystretch}{1.15}
\begin{tabular}{c|cccc|cccc}
\hline
\multirow{2}{*}{\textbf{Method}} 
& \multicolumn{4}{c|}{\textbf{RadioMapSeer}} 
& \multicolumn{4}{c}{\textbf{USC-PL}} \\
\cline{2-9}
& NMSE (dB) & RMSE & MAE & SSIM
& NMSE (dB) & RMSE & MAE & SSIM \\
\hline
ITU-R IMT-2020 
& ~~0.88 & 0.2001 & 0.1469 & 0.5277 
& -4.92 & 0.3743 & 0.3069 & 0.6089 \\

3GPP38.901     
& -2.51  & 0.1355 & 0.0968 & 0.5789 
& -3.57 & 0.4376 & 0.3715 & 0.5072 \\

XGBoost        
& -1.16  & 0.1559 & 0.1217 & 0.4601 
& -15.74 & 0.1046 & 0.0802 & 0.7960 \\

KNN            
& -0.87  & 0.1624 & 0.1183 & 0.5106 
& -14.57 & 0.1182 & 0.0834 & 0.8592 \\

RadioUnet       
& -18.49 & 0.0230 & 0.0120 & 0.9022 
& -25.81 & 0.0324 & 0.0217 & 0.9261 \\

PEFNet         
& -17.89 & 0.0239 & 0.0129 & 0.9006 
& -28.08 & 0.0235 & 0.0157 & 0.9304 \\

ResNet         
& -18.78 & 0.0221 & 0.0124 & 0.8992 
& -28.27 & 0.0236 & 0.0155 & 0.9378 \\

\hline
\textbf{Phy2-ExposNet} 
& \textbf{-19.87} & \textbf{0.0195} & \textbf{0.0102} & \textbf{0.9194}
& \textbf{-30.79} & \textbf{0.0173} & \textbf{0.0112} & \textbf{0.9540} \\
\hline
\end{tabular}
\end{table*}

In contrast, the proposed Phy2-ExposNet in Fig.~\ref{fig:vis}(i) achieves the best agreement with the ground truth. It accurately captures both global attenuation and local structural variations, including sharp shadow boundaries, anisotropic energy distribution, and propagation along streets. This indicates that the proposed physics-informed two stage framework effectively integrates data driven learning with underlying electromagnetic principles, leading to superior reconstruction of EMF exposure maps.

Table~\ref{tab:perf_compare_two_datasets} reports the quantitative comparison on two datasets. It can be observed that empirical models (ITU-R IMT-2020 and 3GPP38.901) and classical machine learning methods (XGBoost and KNN) consistently yield significantly inferior performance across all metrics. Their NMSE and RMSE remain high, and the SSIM values indicate poor reconstruction of spatial propagation structures, suggesting that these approaches are unable to capture complex environment-dependent propagation effects.

We therefore focus on the comparison with deep learning based methods. On the RadioMapSeer dataset, which exhibits relatively regular propagation patterns, all neural models achieve competitive performance. Nevertheless, the proposed Phy2-ExposNet still provides consistent improvements over existing architectures. In particular, it achieves the lowest NMSE and RMSE while maintaining the highest SSIM. Compared with ResNet, Phy2-ExposNet reduces RMSE by approximately $12\%$ and MAE by about $18\%$, while also improving SSIM by around $2\%$, indicating a better balance between numerical accuracy and structural fidelity.

On the more challenging USC-PL dataset, where propagation conditions are more complex and less regular, the performance gap between different methods becomes more pronounced. While baseline models such as RadioUNet and PEFNet still maintain reasonable accuracy, their errors increase noticeably. In contrast, Phy2-ExposNet remains highly stable and achieves the best performance across all metrics.  Even when compared with RadioUNet, Phy2-ExposNet still achieves consistent improvements of about $14\%$ in RMSE.

Overall, these results demonstrate that the proposed method not only improves prediction accuracy on standard datasets, but also generalizes effectively to more challenging environments, highlighting the benefit of integrating physics-informed modeling with data driven learning.

\subsubsection{Ablation Study}
\label{subsubsec:ablation}

To validate the necessity of the proposed two stage architecture and the incorporation of physics-informed losses, we conduct an ablation study by comparing different model configurations. Specifically, we consider three variants: 
(i) a two stage network trained without physics-informed losses, 
(ii) a single stage refinement network trained only with supervised MSE loss, and 
(iii) the proposed Phy2-ExposNet. The predictions are further compared with the ground truth.

\begin{figure*}[t]
    \centering
    \includegraphics[width=\textwidth]{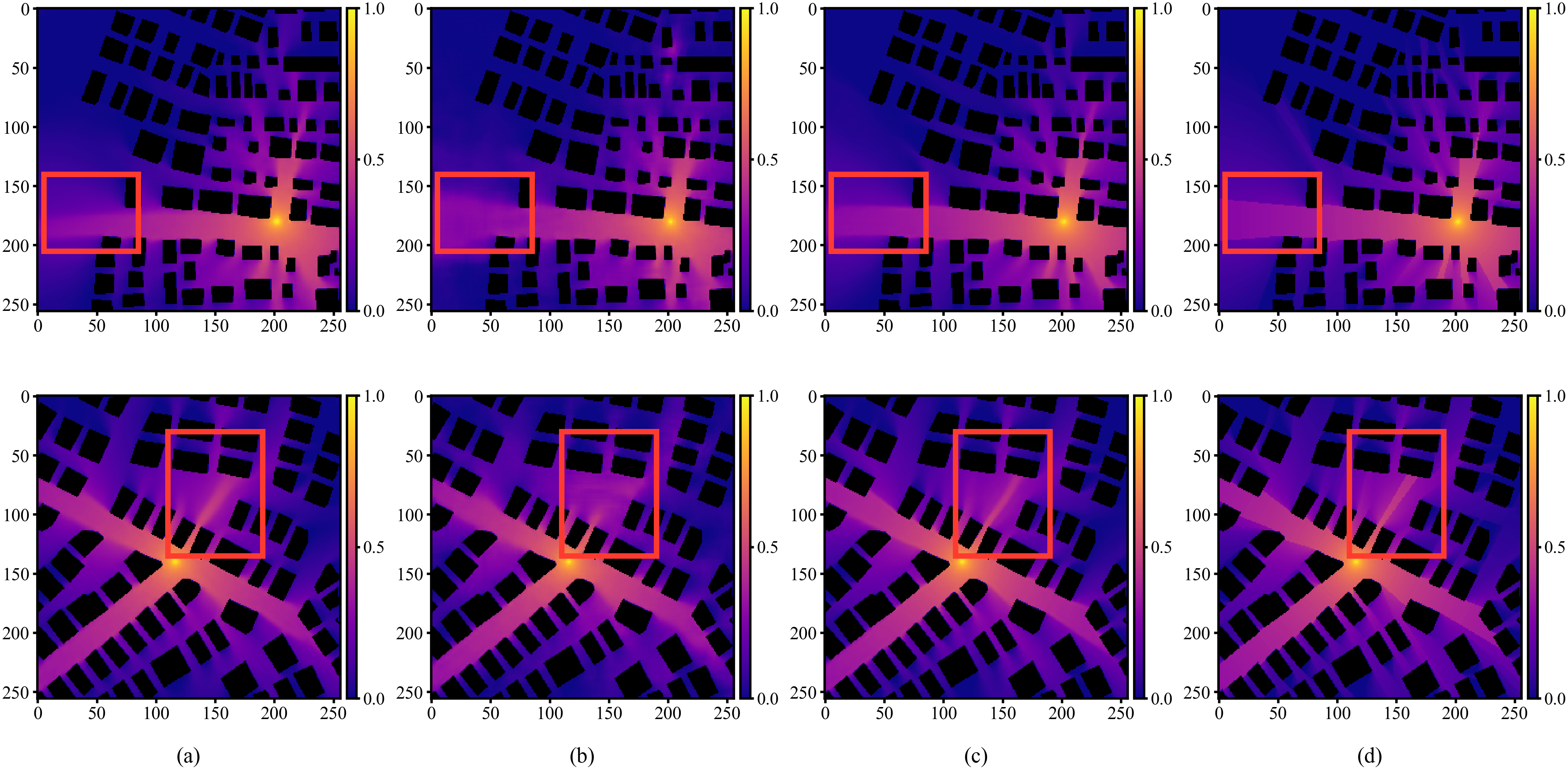}
    \caption{Comparison of ablation models on two test samples. The predicted values are normalized to the range $(0,1)$. From (a) to (d): (a) Two-stage model without physical loss, (b) Stage-2-only model, (c) Phy2-ExposNet, (d) Ground truth.}
    \label{fig:ablation}
\end{figure*}

Fig.~\ref{fig:ablation} presents qualitative results on two test samples. It can be observed that the single stage refinement model, although capable of capturing the overall signal distribution, tends to produce overly smooth predictions and fails to accurately reconstruct electromagnetic spatial variations. Similarly, the two stage model without physics-informed constraints improves the global structure but still exhibits noticeable deviations in complex propagation regions.

In contrast, the proposed Phy2-ExposNet achieves predictions that are visually the closest to the ground truth. In particular, as highlighted in the red boxed regions, the proposed model better captures detailed electromagnetic propagation patterns such as shadowing effects, reflection, and directional signal attenuation. These results indicate that the integration of physics-informed losses effectively guides the model toward physically consistent solutions.

\subsubsection{Accuracy and Efficiency Analysis}
\label{subsubsec:Accuracy and Efficiency }

Table~\ref{tab:efficiency_compare} further compares different models in terms of both prediction accuracy and computational complexity, including parameter count and model size. It can be observed that the proposed Phy2-ExposNet achieves superior accuracy while maintaining a more compact architecture. Compared with representative baselines such as RadioUNet and ResNet, Phy2-ExposNet attains lower prediction error with fewer parameters, demonstrating a more efficient utilization of model capacity.

Such a lightweight yet accurate design is particularly important for practical deployment. In digital twin systems and future wireless communication scenarios, EMF exposure maps are often required to be updated in real time under dynamic environments. Therefore, models with lower computational and memory overhead are more suitable for edge deployment and large-scale inference, enabling faster response and reduced system cost.

To further improve deployment flexibility, we design two lightweight variants, namely Phy2-ExposNet-Light and Phy2-ExposNet-UL (ultra-light). These variants are obtained by progressively reducing the depth and channel capacity of the Stage 1 convolutional network, thereby significantly decreasing the overall parameter count and memory footprint.

Despite this simplification, both variants maintain strong prediction performance. Although a slight increase in error is observed, the degradation remains limited. In particular, the ultra-light version achieves substantial model compression while preserving most of the predictive accuracy. Even with only a fraction of the parameters of the full model, it achieves competitive RMSE and NMSE compared to conventional architectures, and still outperforms RadioUNet in overall accuracy.

\begin{table}[t]
\centering
\caption{Accuracy and complexity comparison of different network variants}
\label{tab:efficiency_compare}
\small

\resizebox{\columnwidth}{!}{
\begin{tabular}{lccccc}
\hline
Method & Params (M) & Size (MB) & NMSE (dB) & RMSE \\
\hline
RadioUNet & 13.27 & 50.64 & -18.49 & 0.0230 \\
PEFNet & 67.41 & 259.12 & -17.89 & 0.0239 \\
ResNet & 12.60 & 50.04 & -18.78 & 0.0221 \\
Phy2-ExposNet & 11.12 & 44.41 & -19.87 & 0.0195 \\
Phy2-ExposNet-Light & 8.67 & 33.07 & -19.45 & 0.0203 \\
Phy2-ExposNet-UL & 3.67 & 14.01 & -19.35 & 0.0206 \\
\hline
\end{tabular}
}
\end{table}

%% ============================================================
%% Conclusion
%% ============================================================

\section{Conclusion}
\label{sec:conclusion}

In this paper, we introduced Phy2-ExposNet, a physics-informed two-stage framework for accurate exposure map estimation. The proposed approach incorporates complementary physical constraints within a hybrid architecture that combines physics-informed field estimation and transformer-driven refinement. Experimental results on two representative datasets show that the model achieves improved estimation accuracy while retaining strong computational efficiency. Furthermore, light and ultra-light variants highlight the practicality of the framework for deployment in resource-constrained scenarios.

Several directions remain open for future work. First, extending Phy2-ExposNet to 3-D environments and higher frequencies is a natural next step.
Second, improving the VIE loss computation for building materials could further tighten the physics constraints in realistic scenarios. Third, incorporating  antenna pattern information and beamforming configurations as additional input features may extend the framework's applicability to massive MIMO network planning. 
Finally, data efficient transfer learning strategies built on the physics-informed representation learned by network could reduce the labeled data requirements for new deployment scenarios.

\section*{Acknowledgment}

This work benefited from a government grant managed by ANR agency under the France 2030 program, reference ANR-22-PEFT-0008.

%% ============================================================
%% References
%% ============================================================
\bibliographystyle{unsrt}
\bibliography{references}

%% ============================================================
%% Author Biographies  
%% ============================================================

\end{document}